\documentclass[11pt]{article}
\usepackage{moriond,epsfig}

\def\Journal#1#2#3#4{{#1} {\bf #2}, #3 (#4)}

\def\PLB{{\em Phys. Lett.}  B}
\def\PRL{\em Phys. Rev. Lett.}
\def\PRD{{\em Phys. Rev.} D}

\def\be{\begin{equation}}
\def\ee{\end{equation}}
\def\bea{\begin{eqnarray}}
\def\eea{\end{eqnarray}}

\begin{document}
\vspace*{4cm}
\title{MEASUREMENT OF THE GLUON POLARISATION AT COMPASS}

\author{G. BRONA, on behalf of the COMPASS collaboration }

\address{University of Warsaw, Faculty of Physics, 00--681 Warsaw, Poland}

\maketitle\abstracts{
COMPASS measurements of the gluon polarisation in nucleon, $\Delta G/G$, are reviewed. Two different approaches based on tagging the Photon Gluon Fusion process are described. They rely on the open charm meson or high--$\mathrm{p}_\mathrm{T}$ hadron pairs detection. The obtained results are: $-0.57 \pm 0.41 \mathrm{\ (stat.)} \pm 0.17 \mathrm{\ (syst.)}$ for the open charm, $0.06 \pm 0.31 \mathrm{\ (stat.)} \pm 0.06 \mathrm{\ (syst.)}$ and $0.016 \pm 0.058 \mathrm{\ (stat.)} \pm 0.055 \mathrm{\ (syst.)}$ for high--$\mathrm{p}_\mathrm{T}$ for $Q^2<1\mathrm{\ (GeV/c)}^2$ and $Q^2>1\mathrm{\ (GeV/c)}^2$ regimes, respectively.}

\section{Introduction} 
In the framework of QCD the nucleon spin can be decomposed into four contributions: from quarks -- $\Delta \Sigma$, from gluons -- $\Delta G$ and from angular momenta of quarks and gluons -- $L_{q}$ and $L_{g}$:   
\begin{equation}\label{deco}
\frac{1}{2}=\frac{1}{2}\Delta\Sigma(\mu^2)+\Delta G(\mu^2)+L_{q}(\mu^2)+L_{g}(\mu^2)
\end{equation}
where $\mu^2$ is a scale at which the nucleon is probed. $\Delta \Sigma$ is determined precisely in a QCD fit to the $g_1$ structure function data and is 0.30$\pm$0.01(stat.)$\pm$0.02(evol.) at 3 $\mathrm{(GeV/c)}^2$ \cite{pap_fit}. This method also gives $\Delta G/G$ albeit with large uncertainty due to the limited kinematical range of the $g_1$ measurements. One of the goals of COMPASS is a determination of the $\Delta G/G$ quantity. The method relies on tagging the Photon Gluon Fusion (PGF) process with high--$\mathrm{p}_\mathrm{T}$ hadron pairs, a channel studied also by HERMES \cite{hermes1,hermes2} and SMC \cite{smc}. In COMPASS also a direct channel based on the PGF tagging with the open charm meson production and decay is used. 

The COmmon Muon and Proton Apparatus for Structure and Spectroscopy (COMPASS) is a two-stage magnetic spectrometer located at CERN SPS muon beam line which delivers 160 GeV/c positive muons of intensity  $2\cdot10^8$ particles per 16.8s SPS cycle. Muons in the beam are naturally polarised. The polarisation is -0.76 (for 2002 and 2003) and -0.81 (for 2004 data taking period). The COMPASS target is composed of two solid state $^6$LiD cells, each 60 cm long and 3 cm in diameter. The spins of the deuterons in the cells are polarised in the opposite directions, parallel and anti--parallel to the beam polarisation. The $^6$Li basically consists of a deuteron plus $^4$He core; a dilution factor of about 0.4 is obtained for the target material. The average polarisation of the target deuterons is $50\%$. Directions of polarisation are flipped every 8 hours. The twin--target and the spin reversals are needed to cancel false asymmetries originating from different spectrometer acceptances for the two target halves and from the time variations of the beam flux. Particles produced in the interactions are traced and identified in two spectrometers equipped with tracking and identification detectors (including RICH), magnets and calorimetry\cite{pap_spectr}.

\section{Determination of $\Delta G/G$}
 
\noindent The analysis is based on the measurement of the cross sections asymmetry of the Photon Gluon Fusion (PGF) interactions with different relative spin orientations of the projectile and of the target nucleon. This asymmetry, $A^{\gamma^* N}$, is coupled to $\Delta G/G$ \textit{via}:
\begin{equation}
A^{\gamma^* N} = R_{PGF}\hat{a}_{LL}\frac{\Delta G}{G}+A_{BG}
\end{equation}
where $\hat{a}_{LL}$ is the partonic asymmetry, $R_{PGF}$ is the fraction of PGF events in the selected sample and $A_{BG}$ is the background asymmetry. Two methods to extract $\Delta G$ are discussed below.

\subsection{The open charm method}

The mass of a charm quark is much larger that of $u$, $d$ and $s$ quarks. The intrinsic charm content of the nucleon at COMPASS kinematics is negligible. Also the production of the charm quarks in the fragmentation process is highly suppressed. The only reaction that has a significant contribution to the charm production is PGF process. The charmed quarks fragment subsequently into charmed hadrons. The detection and identification of them provides a clear tag for the PGF. The studies rely on the detection of $D^0$ mesons in their $\pi K$ decay channel. The mesons are reconstructed by pairing each two charged tracks from a given event and calculating the invariant mass of the system. The signal-to-background ratio is small, of the order of 1:10. Therefore, a second, more exclusive decay channel is studied: $D^{*}\to D^{0}\pi \to K \pi \pi$. In this channel a cut on the $D^*$ mass is imposed and the signal-to-background ratio is increased to approximately 1:1.

Each open charm event is characterised by different value of $\hat{a}_{LL}$. The $R_{PGF}$, corresponding in the open charm channel to the signal-to-background ratio, is a function of the $K\pi$ invariant mass. Therefore a weighted method of the $\Delta G/G$ extraction is applied. Each event is weighted with its $\hat{a}_{LL}$ and $R_{PGF}$. This requires knowledge of the $\hat{a}_{LL}$ on an event-by-event bases. As in the analysis only one charmed meson is required, the reaction kinematics is unknown and the $\hat{a}_{LL}$ cannot be calculated. Thus the parameterisation based on the measured kinematical variables is introduced providing the estimation of the $\hat{a}_{LL}$ for each event. The parameterisation is obtained using Neural Networks trained on the sample prepared with AROMA generator in LO QCD. The correlation between the generated and reconstructed $\hat{a}_{LL}$ is 82$\%$.

The major contributions to the systematic error are listed in Tab~\ref{gbrona:table1}. The background asymmetry is checked in the signal sidebands, in the wrong charge combinations and by simultaneous fitting the signal and the background asymmetries. The data is divided into subsamples, each containing events recorded in approximate a week long intervals. For each the $\Delta G/G$ is calculated. Dispersion of the values is used to estimate the false asymmetries arising from the detector instabilities. The parameter sets in AROMA are varied and a number of $\hat{a}_{LL}$ parameterisations is prepared and used to check the stability of the obtained $\Delta G/G$. Around 300 different fits are used for fitting the signal and the background and the calculation of the $R_{PGF}$.        

A preliminary result from the open charm analysis for 2002-2004 data is:
\begin{equation}
\langle \frac{\Delta G}{G} \rangle = -0.57 \pm 0.41 \mathrm{\ (stat.)} \pm 0.17 \mathrm{\ (syst.)} 
\end{equation}
The average fraction of the nucleon momentum carried by the gluon, $x_g$, for the selected sample is 0.15 with RMS of 0.08. The $\mu^2$ is given by the mass of the charm quark and the $D^0$ meson transverse momentum with respect to photon: $\mu^2=4(m_c^2+p_t^2)$ and equal 13$\mathrm{\ (GeV/c)}^2$. 

\begin{figure}
\epsfig{figure=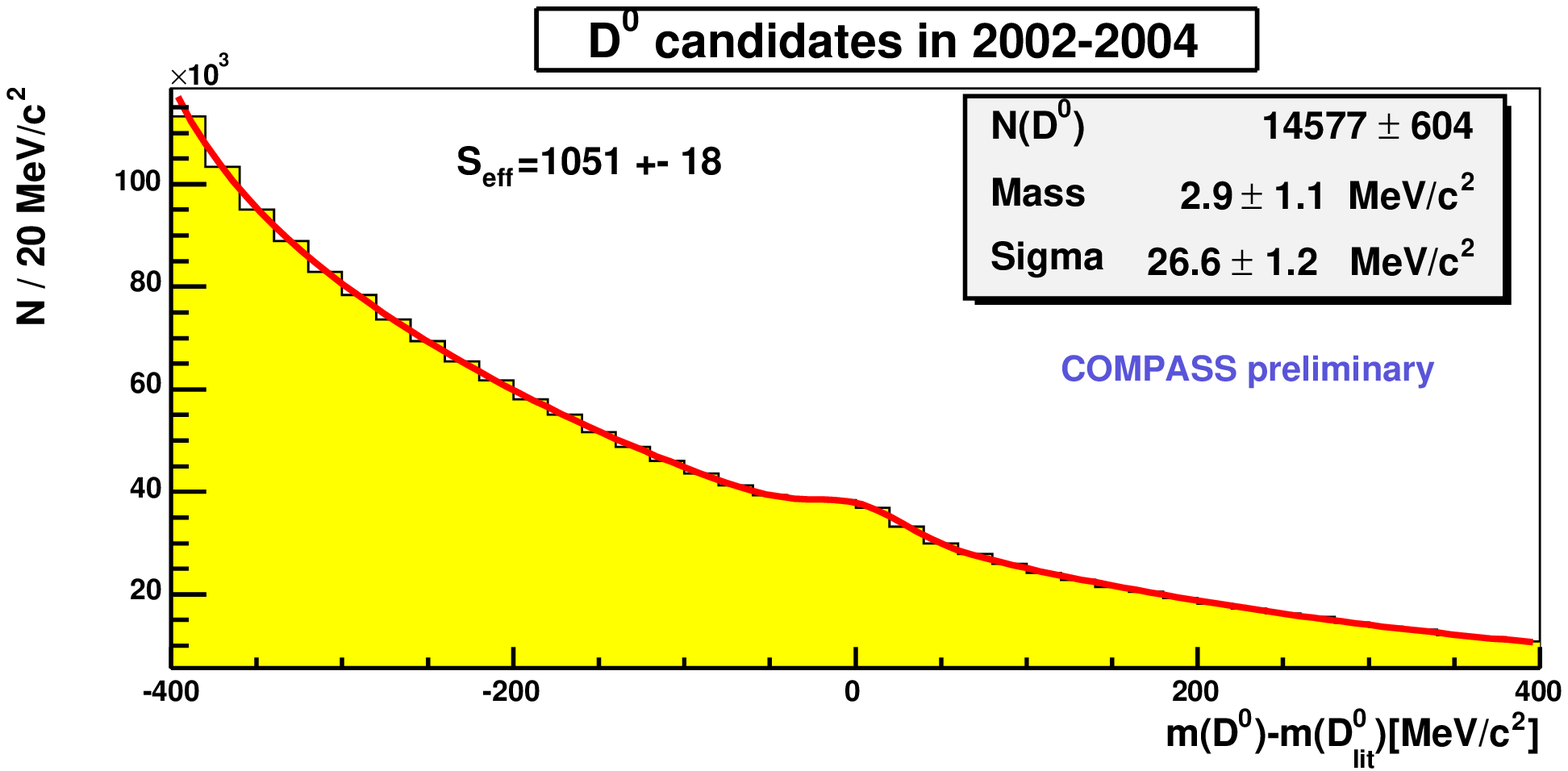,height=1.5in}
\epsfig{figure=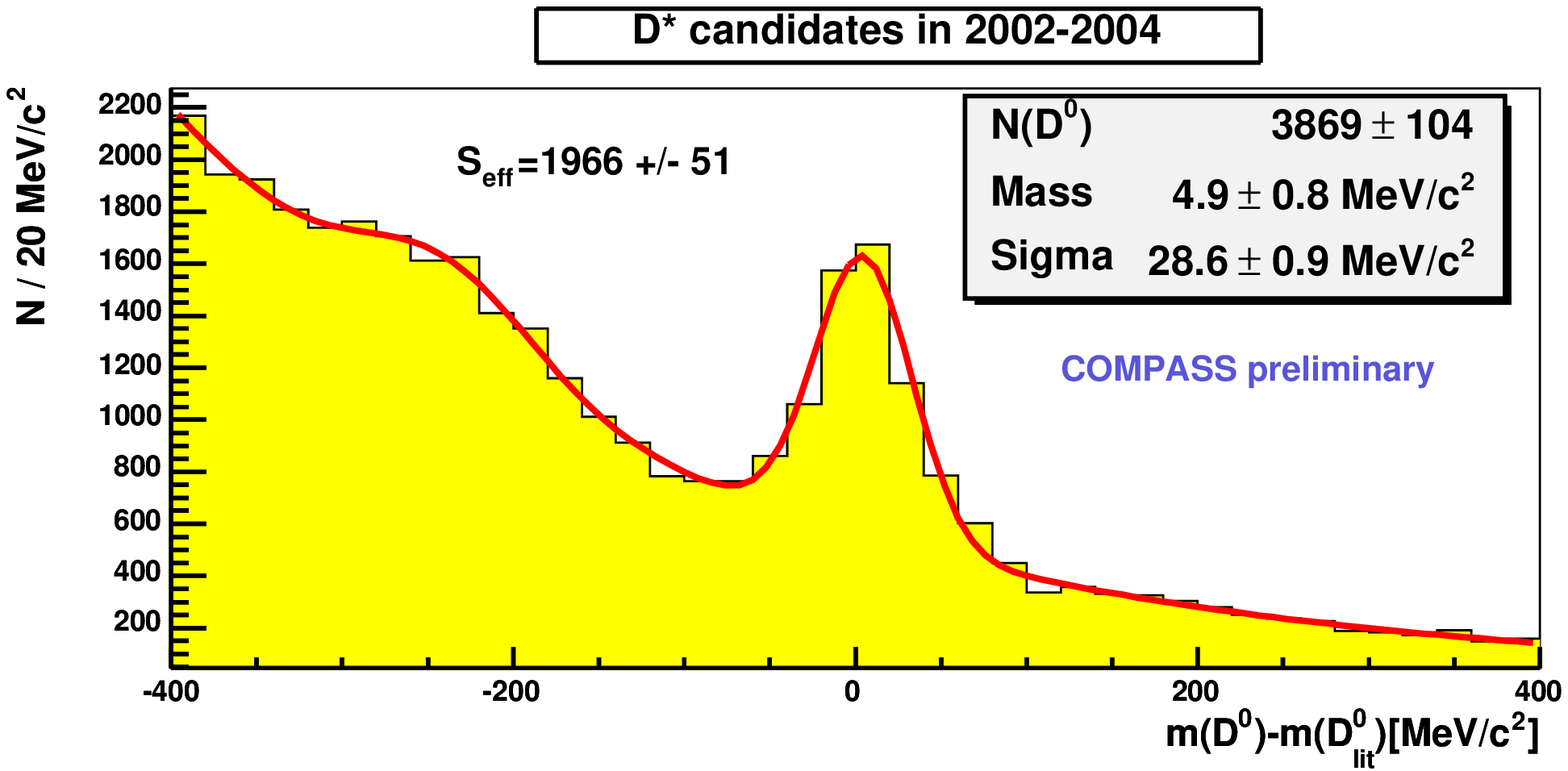,height=1.5in}
\caption{Invariant mass of the $\pi K$ pairs coming from $D^0$ untagged and tagged with $D^*$ samples.}
\end{figure}

\begin{table}[t]
\caption{Major contributions to the systematic uncertainty. They are estimated independently for both channels and found to be equal. The contribution from the uncertainties of the beam and target polarisations, dilution factor and the correlation between the signal strength and the $\hat{a}_{LL}$ added in quadratures are presented as ``Other".\label{gbrona:table1}}
\vspace{0.4cm}
\begin{center}
\begin{tabular}{|l|c|l|c|}
\hline
Background asymmetry & 0.07 & Fitting & 0.09\\
False asymmetries & 0.10 & Other & 0.07 \\
Parameters of AROMA & 0.05 & & \\
\hline
\end{tabular}
\end{center}
\end{table}

\subsection{The high--$\mathrm{p}_\mathrm{T}$ method}

The high--$\mathrm{p}_\mathrm{T}$ method relies on a sample of PGF events with light quark pair production. With the selection of two high--$\mathrm{p}_\mathrm{T}$ hadrons a fraction of PGF events is enhanced and reaches $30\%$. The high--$\mathrm{p}_\mathrm{T}$ analysis is performed in two kinematical regimes:  $Q^2>1\mathrm{\ (GeV/c)}^2$ and $Q^2<1\mathrm{\ (GeV/c)}^2$. 

For the high--$\mathrm{p}_\mathrm{T}$ analysis with $Q^2>1\mathrm{\ (GeV/c)}^2$ the sample contains a large fraction of the Leading Order ($\gamma q \to q$) and QCD-Compton interactions ($\gamma q \to gq$) apart from the PGF. The fraction of the resolved photon processes is assumed to be negligible. The cut $x<0.05$ suppresses the Leading Order and QCD-Compton contributions. Thus: 
\begin{equation}
A^{\gamma^*N \to hh}=R_{PGF}\langle \hat{a}_{LL}^{PGF}\rangle \frac {\Delta G}{G}
\end{equation}
The ratio $R_{PGF}$ is estimated from MC simulation using the LEPTO 6.5.1 generator.

About $90\%$ of events containing two high--$\mathrm{p}_\mathrm{T}$ hadrons have $Q^2<1\mathrm{\ (GeV/c)}^2$. Here apart from the PGF, Leading Order and QCD-Compton the resolved photon processes have to be taken into account. They contribute to more than $50\%$ of interactions. Three channels involving resolved photon dominate: $qq^{\gamma} \to qq$, $qg^{\gamma} \to qg$, $gg^{\gamma} \to gg$. The total asymmetry is:
\pagebreak
\begin{eqnarray}
A^{\gamma^*N \to hh}=\Big[R_{PGF}\langle \hat{a}_{LL}^{PGF}\rangle + R_{qg}\langle \hat{a}_{LL}^{gg}\rangle\Big(\frac{\Delta G}{G}\Big)^{\gamma}\Big]\frac {\Delta G}{G} + \ \ \  \ \ \  \ \ \ \ \ \ \ \ \ \ \ \ \ \ \ \ \ \ \ \ \ \ \ \ \ \ \ \nonumber\\
+\Big[R_{LO} \langle \hat{a}_{LL}^{LO}\rangle + R_{QCDC} \langle \hat{a}_{LL}^{QCDC}\rangle + R_{qq}\langle \hat{a}_{LL}^{qq}\rangle\Big(\frac{\Delta q}{q}\Big)^{\gamma} +  R_{qg}\langle \hat{a}_{LL}^{qg}\rangle \Big(\frac{\Delta G}{G}\Big)^{\gamma} \Big]\frac{\Delta q}{q} \ \ \ \ \ \ 
\end{eqnarray} 
where the superscript $\gamma$ denotes the parton distributions describing the resolved photon structure. The fractions of the processes as well as the average $\hat{a}_{LL}$ for each of them are estimated with a MC simulation using PYTHIA 6.2.  The large number of parameters not measured directly but estimated from the simulation results in a model dependence of the resulting $\Delta G/G$ value. This dependence is encompassed in a systematic error (see also \cite{pap2}). 

The preliminary results from the high--$\mathrm{p}_\mathrm{T}$ analysis from the 2002-2003 data at $Q^2>1\mathrm{\ (GeV/c)}^2$ and for 2002-2004 data at $Q^2<1\mathrm{\ (GeV/c)}^2$ are:
\begin{equation}
\langle \frac{\Delta G}{G} \rangle = 0.06 \pm 0.31 \mathrm{\ (stat.)} \pm 0.06 \mathrm{\ (syst.)} 
\end{equation}
\begin{equation}
\langle \frac{\Delta G}{G} \rangle = 0.016 \pm 0.058 \mathrm{\ (stat.)} \pm 0.055 \mathrm{\ (syst.)} 
\end{equation}
The average $x_g$ for the first one is 0.13 with an RMS of 0.08 while for the second one it is $0.095^{+0.08}_{-0.04}$. The scale $\mu^2$, given by the transverse momentum of the outgoing partons with respect to photon, is 3$\mathrm{\ (GeV/c)}^2$ for both sets.

\begin{figure}
\begin{center}
\epsfig{figure=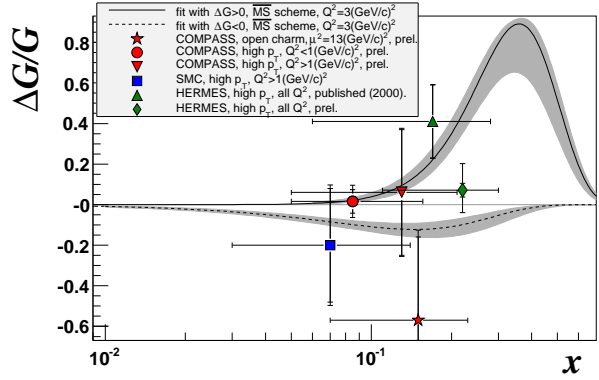,height=2in, bbllx=0pt,bblly=10pt,bburx=560pt,bbury=360pt,clip=}
\end{center}
\caption{Gluon polarisation $\Delta G/G(x_g)$ at $\mu^2=$3 $\mathrm{GeV}^2$ obtained from NLO QCD analysis (curves) and from measurements (points).\label{gbrona:plot2}}
\end{figure}

In Fig. \ref{gbrona:plot2} these direct measurements are compared to parameterisation of $\Delta G(x_g)$ obtained in a NLO fit to the $g_1$ data including the new deuteron results \cite{pap_fit}. Two solutions for $\Delta G(x_g)$ were found, one positive and one negative, both resulting in small $|\Delta G|\sim0.2-0.3$ for $\mu^2=$3 $\mathrm{(GeV/c)}^2$. The error band corresponds to the change of $\chi^2$ by unity. With the present precision measurements cannot distinguish between two possible scenarios. However they are in line with a small value of $\Delta G$.   

\end{document}